\documentclass[aip,pof,preprint]{revtex4-1}
\usepackage[T1]{fontenc}
\usepackage[utf8]{inputenc}
\usepackage{verbatim}
\usepackage{amsmath}
\usepackage{graphicx}
\usepackage{hyperref}
\usepackage{psfrag}
\usepackage{stmaryrd}
\usepackage{amssymb}
\usepackage{wasysym}
\usepackage{hyperref}
\usepackage{ulem}
\usepackage[dvips,usenames,dvipsnames]{color}

\def \degre {$^\mathrm{o}$}

\begin{document}

\title{Experimental observation of a strong mean flow induced by internal gravity waves}

\author{Guilhem Bordes}
\email{guilhem.bordes@ens-lyon.fr}
\author{Antoine Venaille}
\email{antoine.venaille@ens-lyon.fr}
\author{Sylvain Joubaud}
\email{sylvain.joubaud@ens-lyon.fr}
\author{Philippe Odier}
\email{philippe.odier@ens-lyon.fr}
\author{Thierry Dauxois}
\email{thierry.dauxois@ens-lyon.fr}

\affiliation{$^1$Laboratoire de Physique de l'\'Ecole Normale Sup\'erieure de Lyon, CNRS and Universit\'e de Lyon, 46 All\'ee d'Italie, 69007 Lyon, France}

\date{\today}

\begin{abstract}
We report the experimental observation of a robust horizontal mean flow induced by internal gravity waves. A wave beam is forced at the lateral boundary of a tank filled with a linearly stratified fluid initially at rest. After a transient regime, a strong jet appears in the wave beam, with horizontal recirculations outside the wave beam. {Using multiple scale analysis,} we present a simple physical mechanism predicting the growth rate of the mean flow and its initial spatial structure.  We find  good agreement with experimental results. {These results show that a mean flow with non-zero vertical vorticity can be generated by Reynolds stresses if the wave fulfils two conditions: i) the wave amplitude must vary along its propagation direction, which is the case in the presence of viscosity. ii) the wave amplitude must vary in the lateral direction, which is the case when the wave generator is localized in space.}
\end{abstract}

\maketitle

\paragraph*{Introduction.}

Stratified fluids support the existence of anisotropic dispersive waves, called internal gravity waves, which play a major role in astrophysical and geophysical fluid dynamics~\cite{Lighthill1978,Pedlosky1987}. Recent technical advances allowing for  accurate visualization~\cite{Fincham00,Sutherland1999} and  well controlled wave generation~\cite{Gostiaux2007,Mercier2010} in laboratory experiments have provided a renewal of interest in this field~\cite{SutherlandBook}. Previous laboratory experiments focused mostly on propagative wave beams in narrow tanks~\cite{Gostiaux2007,Mercier2010,ManiTomPRL,GostiauxDauxois2007} or propagative vertical modes~\cite{Peacock2009,Joubaud2011}. Here we consider the case of a propagative wave beam in a wide tank, which remains largely unexplored, despite its physical importance. 

A central aspect of wave dynamics is the possible generation of a mean flow due to nonlinearities involving one or several wave beams. These phenomena have important consequences for geophysical flow modeling, since they imply backward energy transfers, or large scale transport properties induced by small scale motions{, see e.g. Ref~\cite{VallisBook,BuhlerBook}}. Among all waves, internal waves are very peculiar because of the specific and unusual nature of nonlinearity. For instance, it has been reported~\cite{DauxoisYoung,Akylas} that in some important cases, the leading nonlinear term unexpectedly cancels out if one has just one internal wave beam.
 
King, Zhang and Swinney~\cite{king} have recently reported the generation of a mean flow  by nonlinearities in the presence of internal gravity waves. However, the structure of the observed mean flow has not been explained, and the underlying mechanism of generation has not yet been proposed. {Lighthill~\cite{Lighthill1978} noticed that either acoustic waves or internal gravity waves can generate steady streaming when their amplitude is attenuated by viscous effects. However, he did not performed explicit computation of the induced mean flow.}  Here we  {fill this gap by performing multiple scale analysis. We provide predictions for the spatial structure and temporal evolution of the mean flow which are} supported by strong experimental evidences.  

There have been  {other} theoretical studies of internal gravity waves-mean flow interactions, see e.g. Refs.~\cite{Bretherton1969,Lelong1991,Akylas}, but none of them considered the case of propagative waves with a slowly varying amplitude in three dimensions. We will show that in our experiments, both viscous attenuation and lateral variations of the wave beam amplitude play a key role in the generation of the observed mean flow. 
 
The paper is organized as follows. We  first present the experimental setup. Then we provide detailed observations of the wave field and of the mean flow. We finally propose a mechanism to deduce the spatial structure  of the mean flow and the  temporal evolution of its amplitude  from the measurements of the wave field.
 
\paragraph*{Experimental setup.} 

\begin{figure}[h]
\includegraphics[height=5.5cm]{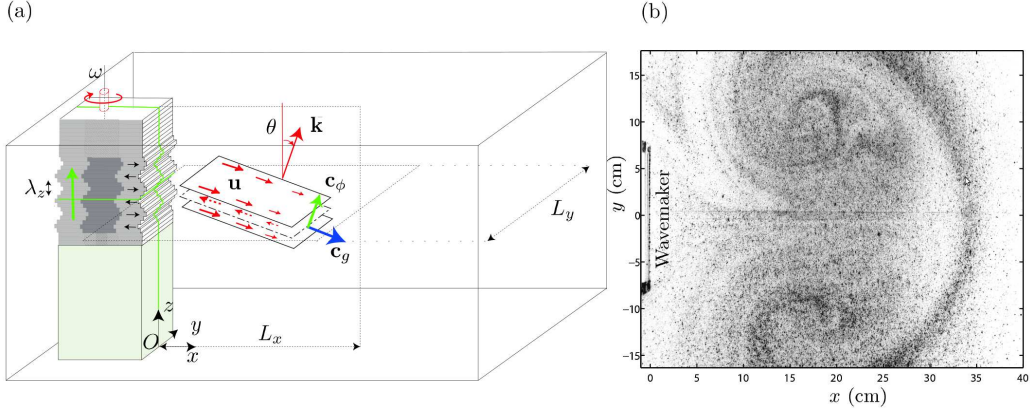}
\caption{(Color online) (a) Schematic representation of the experimental set-up. The wave generator is placed on one side of the tank, defining the origin of the spatial coordinates. The excited plane internal wave has a frequency $\omega$, an upward phase velocity and propagates with an angle $\theta = \sin^{-1}(\omega / N)$. (b) Top view of the particle flow in the horizontal plane $z=21.6$~cm.}
\label{fig:experimental_setup}
\end{figure}

We consider a 120~cm long, 80~cm wide and 42.5~cm deep wave tank, filled with 35~cm of salt water, see Fig.~\ref{fig:experimental_setup}(a). The fluid is linearly stratified in density with a Brunt-V\"ais\"al\"a frequency $N=\sqrt{-(g/\rho)\partial_z\rho}$, where $g$ is the local gravity, $\rho$ the density of the fluid and $z$ the vertical coordinate.  
An internal wave generator is placed on one side of the tank, see Refs.~\cite{Gostiaux2007,Mercier2010} for its full presentation and characterization; it is made of a series of 18 rectangular plates stacked around a helical camshaft. The plates,  which are  $14$~cm wide in the $y$-direction, oscillate back and forth along the longitudinal horizontal coordinate~$x$. The phase shifts between successive cams are chosen in order to form a sinusoidal profile at the surface of the wave generator. The rotation of the camshaft at a frequency $\omega \leq N$ generates a moving boundary condition with an upward or downward phase velocity depending on  the sign of the rotation of the helical camshaft. The displacement profile is $X_0(t,z)=x_0\,\sin(\omega t - m z)$, with a vertical wavelength $\lambda_z=2\pi/m = 3.8$~cm and an amplitude $x_0=0.5$~cm or $x_0=1$~cm.

In the experiments, the front face of the wave generator is located at $x=0$, centered at $y=0,z=15.8$~cm. The  wave beam is $L_{wb}=14$~cm wide,  $11.4$~cm high, corresponding to three wavelengths. The wave generator is only forcing the $x$-component of the internal wave, while the $z$ component is found to adjust according to the internal wave structure. The propagation angle~$\theta$ of the internal wave is varied by changing the rotation rate of the wave generator motor, while keeping the Brunt-V\"ais\"al\"a frequency constant between each experiment, namely $N=0.85$~rad~s$^{-1}$. 
Importantly, this experimental set-up leads to a wave amplitude that depends on the frequency~\cite{Mercier2010}. Moreover,
the axis of the wave generator camshaft staying always vertical, the efficiency of the forcing depends significantly on the projection of the plate motion on the direction of propagation. The wave generator frequency being varied in the range $\omega=0.26\,N$ to $0.50\,N$,  corresponding to an angle of propagation~$\theta$ from $15$\degre~to $30$\degre,  the amplitude of the wave is measured experimentally.

Velocity fields are obtained using a 2D particle image velocimetry (PIV) system~\cite{Fincham00}. The flow is seeded with 10~$\mu$m tracer particles, and illuminated by a 532~nm 2W-continuous laser, shaped into a vertical or horizontal sheet. Respectively, a vertical 35$\times$43~cm$^2$ or horizontal 33$\times$43~cm$^2$ field of view is acquired by a 8-bit 1024$\times$1024 pixels camera. For each wave generator frequency, a set of 600 to 1600 images is recorded, at a frequency of 0.38 to  1.25 Hz, representing 10 images per wave generator period. PIV computations are performed over successive images, on 21$\times$21 pixels interrogation windows with 50\% overlap. The spatial resolution is approximately  $25\times25$ px/cm$^2$. A snapshot of the particle flow is presented in Fig.~\ref{fig:experimental_setup}(b). 
%After some times, a horizontal large scale dipole appears in the fluid.

\begin{figure}[t]
\includegraphics[width=0.8\linewidth]{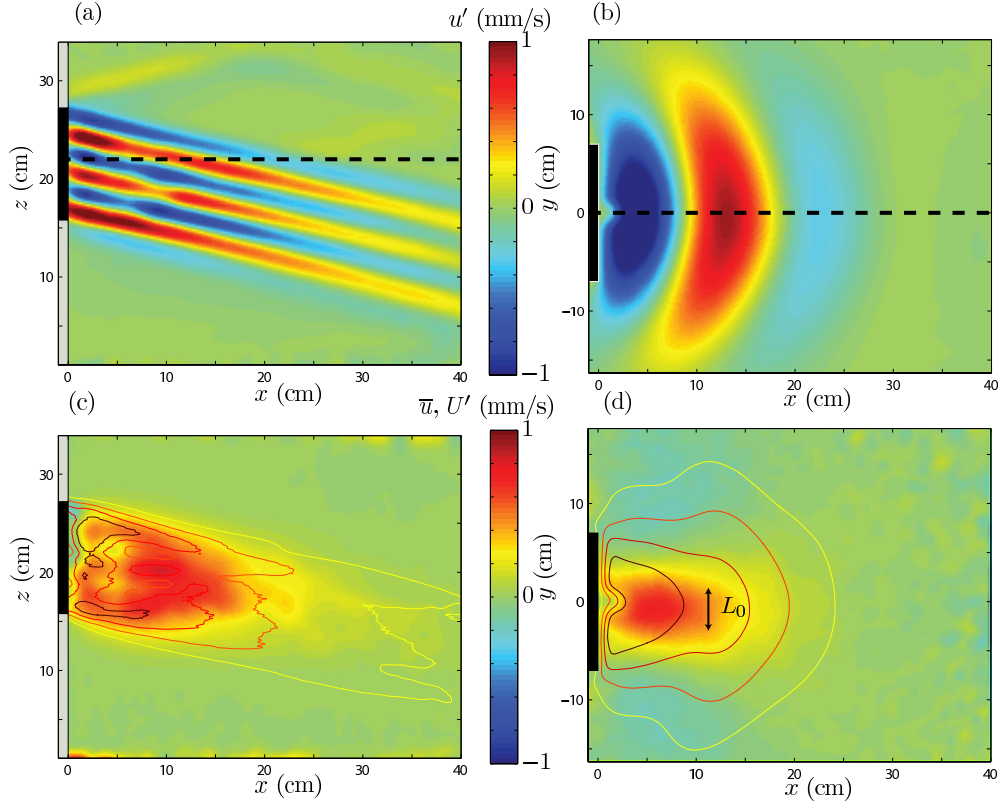} 
\caption{(Color online) (a) and (b): Experimental wave field, $u'$, obtained by filtering the horizontal velocity field at frequency $\omega$. (c) and (d): Experimental mean flow, $\overline{u}$, obtained by low-pass filtering the horizontal velocity field. The contours represent the amplitude of the wave horizontal velocity field, $U'$. The left panels (a) and (c) present the side view and the right panels (b) and (d) present the top view. All pictures were obtained for $\omega/N=0.26$ and a 1~cm eccentricity for the cams.
The wave generator is represented in grey and the moving plates in black. The dashed line in (a) (respectively (b)) indicates the field of view of (b) (respectively (a)).}
 \label{ux_15d_1cm}
\end{figure}

\paragraph*{Experimental results.}

Selective Fourier filtering of the measured horizontal velocity field allows {to distinguish the ``wave'' part of the flow denoted by $(u^{\prime},v^{\prime},w^{\prime})$ from the ``mean'' part of the flow denoted by $(\overline{u},\overline{v},\overline{w})$.}

The wave {part} of the horizontal velocity field is obtained using a bandpass filter centered on {the wave generator frequency} $\omega$, with a width of $0.014\,\omega$ and is presented in Fig.~\ref{ux_15d_1cm}(a) and (b).  Figure~\ref{ux_15d_1cm}(a) shows a vertical slice taken at the center of the wave generator ($y=0$).  We notice the transverse extension over three wavelengths, the amplitude decay in the $x$-direction. Figure~\ref{ux_15d_1cm}(b) presents a horizontal slice located around the mid-depth of the wave generator showing the wave amplitude variations in the $y$-direction. The above experimental observations readily suggests that the wave part of the flow  is monochromatic, {propagating at an angle $\theta$ with respect to the horizontal axis and with} an amplitude varying {slowly in space compared to its wavelength $\lambda$}:
\begin{equation}
u^\prime=U^\prime \cos\left(\omega t-\frac{2\pi}{\lambda}\left(x \sin \theta + z \cos \theta \right)\right) \ . \label{eq:imposed_stream_function}
\end{equation}
{The knowledge of $U^\prime$ is sufficient to determine the other components $v^{\prime}, w^{\prime}$ of the wave field oscillating at frequency $\omega$, see Appendix.} 

%%%%%%%%%%%%
% Mean Flow
%%%%%%%%%%%%
Let us now consider the low-pass filtered flow, {denoted as  $(\overline{u},\overline{v},\overline{w})$}, which we call ``mean flow''.  The width of the filter used to extract the mean flow is $0.25\,\omega$. A strong jet going in the outward direction  from the wave generator is observed in Fig.~\ref{ux_15d_1cm}(c) and (d). This structure is initially located close to the wave generator, then grows until it fills the whole plane. We observe that the jet is precisely produced inside the wave beam. Recirculations on the sides of the tank are visible in  Fig.~\ref{fig:experimental_setup}(b) and in Fig.~\ref{ux_15d_1cm}(d) through the blue patches (blue online). These recirculations clearly show that the horizontal mean flow is characterized by non-zero vertical vorticity {defined as} $\overline{\Omega}=\partial_x \overline{v}-\partial_y \overline{u}$. {Another striking observation is that the vertical velocity of the mean flow could not be distinguished from noise in the experiments. Consequently, the vertical component of the mean flow is negligible with respect to the horizontal components: $\overline{w} \ll \overline{u}$.}

\paragraph*{Interpretation.}

{It has long been known that steady streaming associated with non-zero vertical vorticity can be generated by Reynolds stress due to an internal gravity (or acoustic) wave beam if the wave amplitude is attenuated by viscous effects~\cite{Lighthill1978}, but to our knowledge there exists no explicit computation of this phenomenon. Using multiple scale analysis,} it is possible to go further and propose a {prediction for the spatial structure and the temporal evolution of the mean-flow.}

{Computations are presented in the Appendix, where we consider a slightly more idealized configuration than our experimental setting in order to simplify the analytical treatment: the wave generator is assumed to be tilted in the direction of phase velocity and to be of infinite extension in this direction. We also assume that the velocity amplitude imposed by the wave generator varies smoothly from $0$ to $U$ in the $y$-direction over a distance $L_{wb}$. The three independent adimensionalized numbers of the problem are i) the wave Froude number $\text{Fr}_{\lambda}=U/\left(N\lambda\right)$ ii) the ratio $\lambda/L_{\nu}$ between wavelength and the attenuation length scale $L_{\nu}=\left(\lambda/\left(2\pi\right)\right)^{3}N/\nu$ due to viscous damping in the direction of the wave propagation~\cite{Lighthill1978,Mercier2008}, iii) the ratio $\lambda/L_{wb}$ between wavelength and the lateral extension of the wave generator. We consider a fluid in the Boussinesq approximation~\cite{Pedlosky1987} with the scaling $\mbox{Fr}_{\lambda}=\epsilon^{3}$, $\lambda/L_{\nu}=\epsilon/\lambda_{\nu}$ and $\lambda/L_{wb}=\epsilon/\lambda_{y}$, where $\lambda_{\nu}$ and $\lambda_{y}$ are non-dimensional constants of order one and where $\epsilon$ is the small parameter for the multiple scale analysis. In our experiments, these parameters are estimated as $Fr_\lambda \approx 0.2$, $\lambda/L_{\nu}\approx0.2$ and $\lambda/L_{wb}\approx0.2$, so the assumptions underlying the multiple scale analysis are only marginally satisfied. However, this analysis will allow to get physical insight to the problem. The essential point is that nonlinear terms are negligible at lowest order, and that variations of the wave-amplitude in the $y$ and $x$ direction are of the same order of magnitude.}

{A lengthy but straightforward computation presented in the Appendix shows that at order 5 the dynamics of the vertical component of the filtered vorticity is governed by}
\begin{equation}
\partial_{t}\overline{\Omega} = \left(\overline{\partial_{y}\left(w^{\prime} \partial_z u^{\prime} \right)}-\overline{\partial_{x}\left(w^{\prime} \partial_z v^{\prime} \right)}\right) +\nu \Delta \overline{\Omega} \ ,    \label{eq_vort}
\end{equation}
(corresponding to Eq. (\ref{eq:omega_dyn_interm}) of the Appendix). Note that self-advection of the vertical vorticity by the horizontal mean flow remains negligible at this order. We see that nonlinear terms act as a source of vertical vorticity, in the same manner as the Reynolds stress tensor in a turbulent flow acts as a source of turbulent transport. Using the ansatz (\ref{eq:imposed_stream_function}) the nonlinear term of Eq.~(\ref{eq_vort}) can be explicitly computed (see Appendix for more details) and Eq.~(\ref{eq_vort}) reads:
\begin{equation}
\partial_{t} \overline{\Omega} = \frac{\partial_{xy}U^{\prime2}}{\left(2 \cos\theta\right)^2} +\nu \Delta \overline{\Omega} \  .
%\label{eq:vorticity_dynamics}
\label{eq_vort2}
\end{equation}

A first consequence of Eq.~(\ref{eq_vort2}) is that nonlinearities cannot be a source of  vertical vorticity~$\overline{\Omega}$ if the wave field is invariant in the $y$-direction. In our experiments, the wave generator occupies only one part of the tank width, so one might expect horizontal variations of the wave amplitude, which fulfills the necessary condition to observe generation of a horizontal mean flow associated with non-zero vertical vorticity. A second consequence  is that if the wave field is symmetric in the $y$-direction then the source term in Eq.~(\ref{eq_vort2}), resulting from a derivation of the wave field with respect to~$y$, is antisymmetric in the $y$-direction. This is the case in our experiments, because the wave generator is symmetric with respect to~$y$. A third consequence is that variations in the $x$ directions are necessary to produce vertical vorticity. {Such variation is due to viscous attenuation~\cite{Lighthill1978,Mercier2008} and occurs at a scale $L_{\nu}$ introduced previously.} 

Reference~\cite{Akylas} reported that a single wave beam could not generate a mean flow in a two-dimensional (or axi-symmetric) configuration, with invariance in the y-direction. This is consistent with our finding that the generation of a mean flow requires variations of the wave-amplitude in the $y$ direction. Refs.~\cite{Akylas,AkylasB} also showed that a mean flow  can be generated by two interacting wave beams, even if there is no variation of the wave amplitude in the $y$ direction. Again, there is no contradiction with our results, because the mean flow they describe is associated with zero vertical vorticity, due to their invariance in the $y$ direction.

Finally, let us describe the spatial  structure of the nonlinear term of Eq.~(\ref{eq_vort2}) in our experimental setup. {The wave generator induces a symmetric wave field in the $y$-direction, maximum at the origin, satisfying $\left(\partial_{xy} U^{\prime 2}\right) > 0 $ for $y > 0 $ and $\left(\partial_{xy} U^{\prime 2} \right)< 0 $ for $y<0$.} According to Eq.~(\ref{eq_vort2}), we conclude that the source term induces a dipolar vorticity structure associated with a horizontal jet going in the outward direction from the  wave generator.

\begin{figure}[h]
\includegraphics[height=6cm]{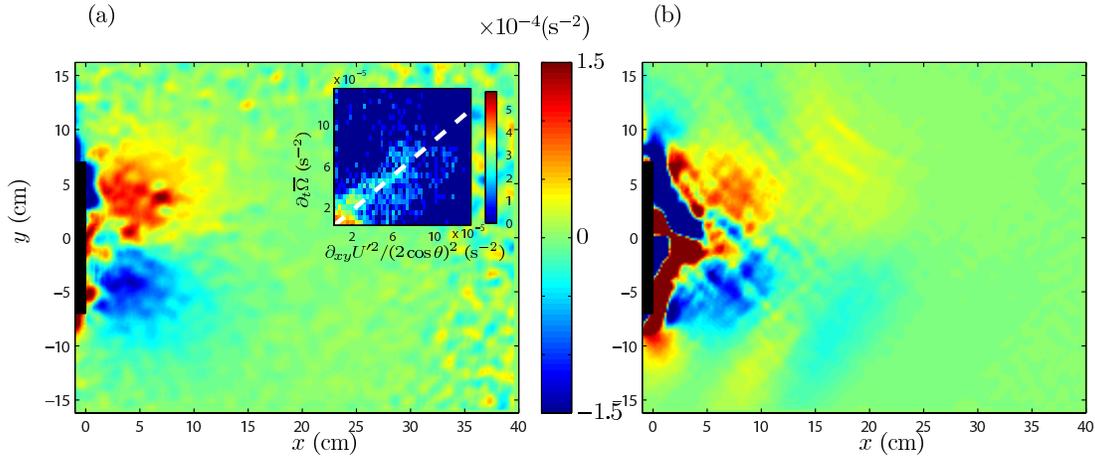}
\caption{(a) and (b) present the top view in the $z=21.6$~cm plane at $t=106$~s of the two important quantities entering in Eq.~(\ref{eq_vort2}):
$\partial_t\overline{\Omega}$ in the left panel and $\partial_{xy}U^{\prime 2}/\left(2\cos\theta\right)^2$ in the right panel. Both pictures were obtained for $\omega/N=0.26$ and a 1 cm eccentricity for the cams. The inset in the left panel presents the 2D-histogram of $\partial_t \overline{\Omega}$ versus $\partial_{xy}U^{\prime 2}/\left(2\cos\theta\right)^2$ to emphasize the correlation between both quantities. The dotted line corresponds to the slope  0.8 ($\pm$0.1) discussed in the text. \label{topview_15d_216}} 
\end{figure}

{Let us now estimate quantitatively this source term. In the experimental results presented here, the wave generator is not tilted with respect to the vertical and not infinite in the direction of phase propagation, but we expect that Eq.~(\ref{eq_vort2}) remains valid if we consider the flow at mid-depth of the wave generator sufficiently far from the lateral boundary.}
%\DEL{ We also performed experiments with a tilted generator and did not find  strong differences provided that the angle of propagation remained sufficiently small.}
%%%%%%%%%%%%%%%%%%%%%%%%%%%%%%%%%%%%%%
% Dipolar structure of the source term.
%%%%%%%%%%%%%%%%%%%%%%%%%%%%%%%%%%%%%%
{Consistently with this remark,} the horizontal slice shown in Fig.~\ref{ux_15d_1cm}(d) is centered on the middle plate of the wave generator (dashed line in Fig.~\ref{ux_15d_1cm}(a)), 
%{in order to be as close as possible to the case of an infinite generator in the direction of the phase velocity. The experimental results are presented for the case of a vertical generator while we assumed in the analytical compuation that the generator was tilted at an angle $\theta$ with respect to the vertical axis. We checked on one case that there were no difference when the generator was tilted provided that the angle of propagation is sufficiently small.} 
%\DEL{$\partial_z \Psi^2$=0 in Eq.~(\ref{eq_vort_stream function}), except at the top and bottom edges of the wave beam (see Fig.~\ref{ux_15d_1cm}(a)). Assuming also that the wave beam is sufficiently damped by viscosity as its top edge crosses the horizontal slice, we will neglect the term $\partial_z \Psi^2$  everywhere.}
%\DEL{Assuming  finally that the wavelength $2\pi/m$ is much smaller than typical spatial variations of~$\Psi^2$, the wave stream function is given by  $\Psi=U^{\prime}/m$, where $U'$ is the amplitude of the horizontal velocity field of the wave, $u^{\prime} = -U^{\prime}(x,y,z)\sin(\omega t -kx -mz)$. Equation~(\ref{eq_vort_stream function}) becomes} 
%\begin{equation}
%\partial_{t}\Omega_{0z}+\mathbf{u}_{0}\cdot\nabla\Omega_{0z}=\frac{1}{2}\partial_{xy}U^{\prime 2} \ +\nu \Delta \Omega_{0z}.\label{eq_vort_simplified}
%\end{equation}
For sufficiently short times, the viscous term of Eq. (\ref{eq_vort2}) can be neglected, and one expects that $\partial_{t}\overline{\Omega} \simeq \left(\partial_{xy}U^{\prime 2}\right)/ \left(2\cos\theta\right)^2$. The experimental determinations of these two terms are compared in Fig.~\ref{topview_15d_216} for $t=106$~s. This time is much smaller than the viscous time ($\sim 1000$~s) described in the last part of this paper. The magnitude of $\partial_t \overline{\Omega}$ (Fig.~\ref{topview_15d_216}(a)) is of the order of the source term  (Fig.~\ref{topview_15d_216}(b)) and the two fields correspond well spatially. The inset in Figure~\ref{topview_15d_216}(a) shows a good spatial correlation between both fields. The slope that relates the two terms is 0.8 $(\pm0.1)$, {consistently with} the theory, which predicts $1$. The {small discrepancy} may be due {the fact the wave generator is not tilted and not infinite in the direction of phase propagation.}

%%%%%%%%%%%%%%%%%%%%%%%%%%%%
%Growth of the jet strength
%%%%%%%%%%%%%%%%%%%%%%%%%%%%
In order to address the temporal evolution of the mean horizontal flow at mid-depth of the wave generator, it is convenient to integrate Eq.~(\ref{eq_vort2}) over a domain defined as a half-plane of Fig.~\ref{topview_15d_216}(b)  (domain  $\mathcal{D}^+= \left[0, L_{x} \right] \times \left[0,  \ L_{y}/2\right]$). One obtains 
\begin{equation}
\partial_t  \iint_{\mathcal{D}^+} \mathrm{d}x \mathrm{d}y\,\overline{\Omega}= \left(\frac{U^{\prime}(0,0,z)}{2 \cos\theta }\right)^2+\nu \iint_{\mathcal{D}^+} \mathrm{d}x \mathrm{d}y \ \Delta \overline{\Omega}.  \label{eq:growth0}
\end{equation}
The integral in the left-hand side of this equation, which we call $I(z,t)$, can be transformed as (if we denote ${\mathcal{C}^+}$ the circulation path surrounding the domain ${\mathcal{D}^+}$)
\begin{equation}
 I(z,t)\equiv \iint_{\mathcal{D}^+}  \mathrm{d}x \mathrm{d}y  \ \overline{\Omega}(x,y,z,t) =
 \oint_{\mathcal{C}^+}\mathrm{d} \mathbf{l}\cdot \overline{\mathbf{u}}
 \approx \int_{0}^{L_{x}}  \mathrm{d} x\ \overline{\mathbf{u}}, \label{eq:jet_strength}
 \end{equation}
since along ${\mathcal{C}^+}$, the velocity is non zero only on the $y$=0 axis. In this form, $I(z,t)$ can be identified as a measure of the jet strength. Equation~(\ref{eq:growth0}) can then be written as
\begin{equation}
\partial_t  I(z, t)= \left(\frac{U^{\prime}(0,0,z)}{2 \cos\theta}\right)^2+\nu \iint_{\mathcal{D}^+} \mathrm{d}x \mathrm{d}y \ \Delta \overline{\Omega}.  \label{eq:growth}
\end{equation}
Finally, defining $L_0$ as the smallest  characteristic scale, i.e. along $y$, of the vorticity $\overline{\Omega}$, the viscous term may be approximated by $-\nu I(z,t) /L_0^2$. The jet strength is then solution of a first order differential equation,
\begin{equation}
\partial_t  I(z, t)= S-\frac{\nu}{L_0^2} I(z,t),  \label{eq:growthbis}
\end{equation}
where $S$ is the source term $S=U^{\prime 2}(0,0,z)/ \left(2 \cos \theta \right)^2$. Equation~(\ref{eq:growthbis}) shows that the jet strength $I(z,t)$ should vary exponentially with time and one gets
\begin{equation}
I(z,t) = \frac{SL_0^2}{\nu}\left(1-e^{-\nu t/L_0^2}\right)\,.  \label{eq:growthsol}
\end{equation}

\begin{figure}[h!]
\begin{center}
\includegraphics[height=6cm]{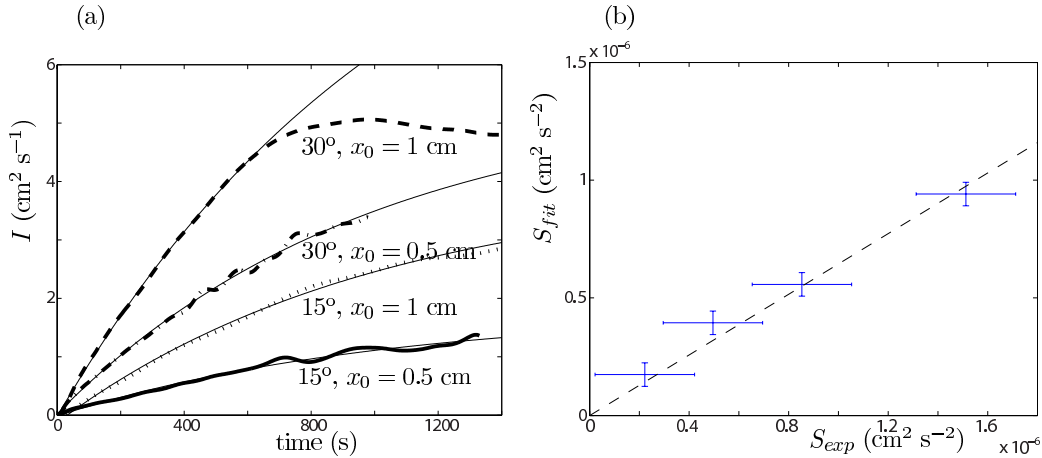} 
\caption{(a) presents the evolution vs time of the quantity  $I(z_0,t)$, with exponential fits. (b) shows the value $S_{\rm fit}$ of the source term estimated from the exponential fits of panel (a) vs the experimental estimation $S_{\rm exp}$ of the same source term.  The dashed line corresponds to $S_{\rm fit}= (0.7\pm0.1)\, S_{\rm exp}$.}
\label{time_evol}
\end{center}
\end{figure}

We observe such an exponential growth in Fig.~\ref{time_evol}(a), which represents the time evolution of the quantity $I(z,t)$ for different values of  $S$ and $z=21.6$~cm. A deviation from the exponential growth  happens at large times when the mean flow reaches the side of the visualization window, in which case the integral $I(z,t)$ saturates since the approximation of Eq.~\ref{eq:jet_strength} is not valid any more. The value of $L_0^2$ is estimated by an exponential fit of the evolution of $I(z,t)$ as a function of time. Remarkably, one  gets the same characteristic scale $L_0 \approx 4$~cm for all experiments, which corresponds to the jet width shown in Fig.~\ref{ux_15d_1cm}(d). We then estimate experimentally the source term in two different ways: the exponential fit ($S_{\rm fit}$) and the measurement of the amplitude of the horizontal velocity field ($S_{\rm exp}=\left(U^\prime(0,0,z)/\left(2\cos\theta\right)\right)^2$). $S_{\rm fit}$ is plotted as a function of $S_{\rm exp}$ in Fig.~\ref{time_evol}(b). As expected, a linear relation between these two estimations is obtained. We find $S_{\rm fit}=(0.7\pm0.1)\,S_{\rm exp}$ in agreement with the spatial correlation observed in the inset of Fig.~\ref{topview_15d_216}(a). However our simple model predicts $S_{\rm fit}=S_{\rm exp}$. {We notice that the smaller the angle, the closer to the theoretical prediction is the experiment. The reason is the the wave generator is not tilted with the vertical axis in the direction of phase propagation, contrary to the assumption done for the multiple scale analysis, so our prediction become less accurate with increasing the angle $\theta$.} 

{The fact that we obtain a fairly good prediction for the the long time evolution of the jet strength may seem surprising, since Eq. (\ref{eq_vort2}) holds only for sufficiently small time. The main reason is that the integration procedure over the half plane would cancel the additional horizontal self-advection term $\overline{\mathbf{u}}_H \cdot \nabla  \overline{\Omega}$ in Eq. (\ref{eq_vort2}), even if it may be locally important. Indeed, considering that the horizontal mean flow is non-divergent, the advection term can be written  ${\nabla \cdot\left(\overline{\Omega} \overline{\mathbf{u}}_H \right)}$, and the surface integral of this term vanishes since $\overline{\Omega}$ is zero on the $Ox$ axis and the filtered horizontal velocity field is zero on the other edges of the domain. Consequently, we expect that Eq. (\ref{eq:growthsol}) remains valid at time larger than those required to fulfill the hypothesis of the multiple scale analysis.} 

\paragraph*{Conclusion.}
We have reported experimental observations of a strong horizontal mean flow with non-zero vertical vorticity when a propagative monochromatic wave is forced on the side of a tank filled with a linearly stratified fluid. We stress here that there is no such mean flow in absence of internal waves propagation, for example when the wave generator is excited at a frequency $\omega$ larger than the Brunt-V\"ais\"al\"a frequency~$N$, as in Ref.\cite{king}. 

The key ingredient for the existence of this mean vertical vorticity with a dipolar structure, associated with a strong horizontal jet flowing outward of the wave generator is the concomitant existence of variations of the wave amplitude in both horizontal directions.
In the transverse direction ($y$), the variations are simply due to the fact that the wave generator is localized in a segment smaller than the tank width. In the longitudinal direction, the variations of the wave amplitude are due to viscous attenuation. This shows the important role played by viscosity in the generation of the mean flow in our experiments. 

{Interestingly, a similar phenomenon as the one described in this paper has been observed in the case of the reflexion of an internal wavebeam over topography}~\cite{Leclair2011}. {Although viscosity is essential for the wave beam attenuation in the experiments, several other physical mechanism may enhance its effect.} {For instance, parametric subharmonic instabilities followed by a direct energy cascade~\cite{Joubaud2011} or wave breaking~\cite{Peacock2009} may lead to a spatially varying wave beam amplitude.} Rather than considering viscous effects, Ref.~\cite{Grisouard2012} models the effect of wave breaking by a linear damping term in the buoyancy equation and uses generalized Lagrangian-mean theory to describe the formation of a strong mean flow. This gives a complementary point of view to the present paper, which is focused on the description of a laminar experiment. We hope to address the effect of wave-breaking in future experimental work.

\acknowledgments
The authors thank M. Lasbleis  for preliminary experiments, {two anonymous reviewers for their input and Nicolas Grisouard for useful comments on a first draft}. ENS Lyon's research work has been supported by the grants ANR-08-BLAN-0113-01 ``PIWO'',  ANR-2011-BS04-006-01 ``ONLITUR'' and CIBLE 2010 from R\'egion Rh\^one-Alpes.

\section*{Appendix: Multiple scale analysis}

%\textcolor{blue}
{
We consider a Boussinesq fluid linearly stratified with buoyancy frequency $N$ and viscosity~$\nu$, and a wave generator oscillating at frequency $\omega = N \sin \theta$ with a wavelength $\lambda$. The wave generator is tilted at an angle $\theta$ with the vertical axis. The amplitude of the horizontal velocity imposed by the wave generator is denoted as $A_0(y)U$. We adimensionalize length, time, velocity and buoyancy $b$  by $\lambda/\left(2\pi\right)$, $1/N$, $U$ and $NU$, respectively. In terms of the velocity components $(u,v,w)$, of the vertical vorticity $\Omega=v_{x}-u_{y}$  and of the buoyancy $b$, the dynamics reads 
%\begin{equation}
%\widetilde{\mathbf{u}}=\frac{\mathbf{u}}{U},\quad\widetilde{\mathbf{x}}=\frac{\mathbf{x}}{\lambda},\quad\widetilde{t}=t%N,\quad\widetilde{b}=\frac{b}{NU},
%\label{eq:adimensionalize}
%\end{equation}
%where $U$ and $\lambda$ are respectively the velocity and the wavelength imposed by the wave generator. 
\begin{equation}
\Omega_{t}+\epsilon^{3}\left(\mathbf{u}_{H}\cdot\nabla_{H}\Omega+\left(\nabla_{H}\cdot\mathbf{u}_{H}\right)\Omega+\partial_{x}\left(w \partial_z v \right)-\partial_{y}\left(w \partial_z u\right)\right)=\epsilon\lambda_{\nu}^{-1}\Delta \Omega,\label{eq:OmegaAD-1}\end{equation}
\begin{equation}
\Delta w_{tt}+\Delta_{H}w=\epsilon\lambda_{\nu}^{-1}\Delta\Delta w_{t}-\epsilon^{3}\left(\partial_{t}\left(\Delta_{H}\left(\mathbf{u}\cdot\nabla w\right)-\partial_{z}\nabla_{H}\left(\mathbf{u}\cdot\nabla\mathbf{u}_{H}\right)\right)+\Delta_{H}\left(\mathbf{u}\cdot\nabla b\right)\right),
\label{eq:WAD}
\end{equation}
\begin{equation}
\nabla_{H}\cdot\mathbf{u}_{H}=-w_{z},
\label{eq:divAD}
\end{equation}
\begin{equation}
b_{t}+\epsilon^{3}\left(\mathbf{u}\cdot\nabla b\right)+w=0,
\label{eq:BAD}
\end{equation}
where the indices $t,x,y,z$ stand for partial derivatives, and where we have considered the scaling proposed in the text for the adimensional parameters ($U/\left(\lambda N\right)=\epsilon^3$, $\nu/\left(\lambda^2 N\right)=\epsilon/\lambda_{\nu}$). We denote $\mathbf{u}_H=\left(u,v,0\right)$ the horizontal velocity field. Since the wave generator oscillates at frequency $\sin\theta,$ we introduce the appropriate coordinates
\begin{equation}
\xi=x\cos\theta-z\sin\theta\quad  \text{and} \quad \eta=x\sin\theta+z\cos\theta\label{eq:new_coordinate}
\end{equation}
(see Fig. 5). The wave generator is located at $\xi=0$ and is of infinite extension in the $\eta$-direction. This amounts to impose the boundary condition
\begin{equation}
w(\xi=0,\eta,y,t)=A_{0}(y)\sin\left(t \sin\theta -\eta \right)\label{eq:boundary_w}
\end{equation}
The initial condition is such that there is no flow in the domain at $t=0$ ($u=v=w=0$ and $\Omega=0$).
\begin{figure}[b]
\begin{centering}
\includegraphics[height=4cm]{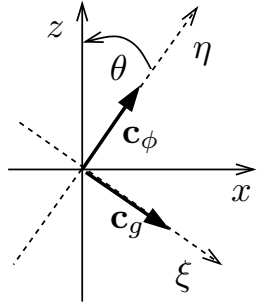}
\end{centering}
\caption{The wave generator is assumed to be of infinite extension in the $\eta$-direction, located at $\xi=0$. Arrows $c_{\phi}$ and $c_{g}$ are phase and
group velocities, respectively.}
\end{figure}
We introduce the rescaled coordinates
\begin{equation}
\eta_{0}=\eta,\quad\xi_{1}=\epsilon\xi,\quad\xi_{2}=\epsilon^{2}\xi\quad y_{1}=\epsilon y,\quad t_{i}=\epsilon^{i}\ \text{with \ensuremath{i=0,1,...}}.
\label{eq:dvpt3}
\end{equation}
%\begin{comment}
%In the rescaled coordinate system, we have
%\begin{equation}
%\Delta\equiv\partial_{\xi\xi}+\partial_{\eta\eta}+\partial_{yy},\quad\Delta_{H}\equiv\cos^{2}\theta\partial_{\xi\xi}+\sin^{2}\theta\partial_{\eta\eta}+2\cos\theta\sin\theta\partial_{\xi\eta}
%\label{eq:Laplacian}
%\end{equation}
%\end{comment}
and 
\begin{equation}
w=w_{0}\left(t_{0},t_{1},t_{2},t_{3},\eta,y_{1},\xi_{1},\xi_{2}\right)+\epsilon w_{1}\left(t_{1},t_{2},t_{3},y_{1},\xi_{1},\xi_{2}\right)+o(\epsilon)\ ,
\label{eq:dvpt1}
\end{equation}
\begin{equation}
u=u_{0}\left(t_{0},t_{1},t_{2},t_{3},\eta,y_{1},\xi_{1},\xi_{2}\right)+\epsilon u_{1}\left(t_{1},t_{2},y_{1},\xi_{1},\xi_{2}\right)+o(\epsilon)
\label{eq:u_dvpt_def}
\end{equation}
\begin{equation}
v=\epsilon v_{1}\left(t_{0},t_{1},t_{2},t_{3},\eta,y_{1},\xi_{1},\xi_{2}\right)+o(\epsilon) \ ,
\label{eq:v_dvpt_def}
\end{equation}
\begin{equation}
\Omega=\epsilon^{2}\Omega_{2}\left(t_{3},y_{1},\xi_{1},\xi_{2}\right)+\epsilon^{4}\Omega_{4}\left(t_{0},t_{1},t_{2},t_{3},\eta,y_{1},\xi_{1},\xi_{2}\right)+o(\epsilon^{4}) \ .
\label{eq:dvpt2}
\end{equation}
 The strategy is the following: computations at order $0,1,2$ allow
to obtain the structure of the wavebeam by solving Eq. (\ref{eq:WAD}).
At these orders, nonlinear terms are negligible and there can consequently
be no source of vertical vorticity. We then consider the dynamics
of the vertical vorticity given by Eq. (\ref{eq:OmegaAD-1}) at order
$3,4,5$ to find the structure of the vortical flow induced by the
wavebeam through Reynolds stresses, considering a temporal filtering
over one wave-period. Our scaling is such that generation of a mean-flow
through Reynolds stresses due to the wave attenuation appears at order
$5$, and requires the knowledge of the wave structure at order $2$.
Note that using the notations introduced in the text, the mean vortical flow is given by $\overline{\Omega}=\overline{\Omega_2}=\Omega_2$ while the wave field is $(u^\prime,v^\prime,w^\prime)=(u_0,\epsilon v_1, w_0)$. Indeed, we can show a posteriori that $\overline{\Omega}_4=0$ at order 5.\\
At zeroth order, we recover from Eq. (\ref{eq:WAD}) the inviscid dispersion
relation, while the boundary condition imposes the wavelength of the wave:
\begin{equation}
w_{0}=W_{0}e^{-i\eta_{0}+i t_0 \sin\theta}+W_{0}^{*}e^{i\eta_{0}-i t_0 \sin\theta}+f_{w},\quad f_{w\eta_{0}} =0\ .
\label{eq:W0_ordre0}
\end{equation}
Equation (\ref{eq:divAD}) gives the corresponding velocity in the
$x$-direction $u_{0\eta_{0}}=w_{0\eta_{0}}/\tan\theta$. Considering order 1 terms in Eq. (\ref{eq:WAD}), we obtain that for $t_{1}>\xi_{1}/\cos\theta$, 
\begin{equation}
W_{0}=\mathcal{A}_{0}\ e^{-\xi_{1}/\left(2\lambda_{\nu}\cos\theta\right)},\quad \text{with} \quad \mathcal{A}_{0}(t_{2},t_{3},y_{1},\xi_{2}=0)=A_{0}(y_{1}),
\label{eq:sol_ordre1_W0_zeta1}
\end{equation}
Using $ v_{1 x_0} =  u_{0 y_1}$ leads to 
\begin{equation}
v_{1}=\frac{i\cos\theta}{\left(\sin\theta\right)^{2}}e^{-\xi_{1}/\left(2\lambda_{\nu}\cos\theta\right)}\left(\mathcal{A}_{0y_{1}}^{*}e^{i\eta_{0}-it_0 \sin\theta}-\mathcal{A}_{0y_{1}}e^{-i\eta_{0}+it_0 \sin \theta}\right)+f_{v},\quad f_{v\eta_{0}}=0\ .
\label{eq:v1-1}
\end{equation}
Considering the order 2 terms in Eq. (\ref{eq:WAD}), one can show
that for $t_{2}>\xi_{2}/\cos\theta$, $w_{0}$ does not depend on
$t_{2}$. At order 3, all nonlinear terms appearing in Eq. (\ref{eq:OmegaAD-1})
for the dynamics of $\Omega$ vanish. At order 4, we obtain
\begin{equation}
\Omega_{4t_{0}}+\left(\partial_{x_{0}}\left(w_{0}\partial_{z_0}v_{1}\right)-\partial_{y_{1}}\left(w_{0} \partial_{z_0} u_{0}\right)\right)=0\ .\label{eq:order4_1}
\end{equation}
%\begin{comment}
%\begin{equation}
%\partial_{x_{0}}\left(w_{0}v_{1z_{0}}\right)-\partial_{y_{1}}\left(w_{0}u_{0z_{0}}\right)=\left(\left(W_{0}V_{1z_{0}}\right)_{x_{0}}-\left(W_{0}V_{1z_{0}}\right)_{y_{1}}\right)e^{2i\sin\theta t_{0}-2i\eta_{0}}+\left(\left(W_{0}^{*}V_{1z_{0}}^{*}\right)_{x_{0}}-\left(W_{0}^{*}U_{0z_{0}}^{*}\right)_{y_{1}}\right)e^{-2i\sin\theta t_{0}+2i\eta_{0}}+...\label{eq:order_4_2}\end{equation}
%\[...\partial_{x_{0}}\left(-iW_{0}V_{1}^{*}\eta_{0z_{0}}+iW_{0}^{*}V_{1}\eta_{0z_{0}}\right)+\partial_{y_{1}}\left(-i\eta_{0z_{0}}W_{0}U_{0}^{*}+i\eta_{0z_{0}}W_{0}^{*}U_{0}\right)\]
%Importantly, both terms on the second line are zero, since $V_{1}W_{0}^{*}-W_{0}V_{1}^{*}=-i\frac{\cos\theta}{\sin^{2}\theta}\left(\mathcal{A}_{0}\mathcal{A}_{0y_{1}}^{*}+\mathcal{A}_{0}\mathcal{A}_{0y_{1}}^{*}\right)$
%does not depend on $x_{0}$, and since $W_{0}^{*}U_{0}-W_{0}U_{0}^{*}=0$. 
%\end{comment}
Using Eqs. (\ref{eq:sol_ordre1_W0_zeta1}-\ref{eq:v1-1}) and $u_{0\eta_{0}}=w_{0\eta_{0}}/\tan\theta$,
one can show that $\partial_{x_{0}}\left(w_{0}\partial_{z_0}v_{1}\right)-\partial_{y_{1}}\left(w_{0} \partial_{z_0} u_{0}\right)$
oscillates at frequency $2\sin\theta$ and that the term independent
of $t_{0}$ vanishes: the nonlinear terms do not induce a steady streaming
at this order.\\
%
%\begin{comment}
%It implies that $\Omega_{4}=Ce^{i\sin\theta t_{0}}+De^{2i\sin\theta t_{0}}+\Omega_{4}$
%with $\Omega_{4t_{0}}=0$.
%The dynamics of vorticity at order 5 reads 
%\begin{equation}
%\Omega_{2t_{3}}+\Omega_{4t_{1}}+\Omega_{5t_{0}}+\partial_{x_{0}}\left(u_{0}\Omega_{2}\right)+\partial_{x_{0}}\left(w_{0}v_{1z_{1}}\right)+\partial_{x_{1}}\left(w_{0}v_{1z_{0}}\right)+\partial_{x_{0}}\left(w_{1}v_{1z_{0}}\right)+...\label{eq:order5_1}\end{equation}
%\[...-\partial_{y_{1}}\left(w_{0}u_{1z_{0}}\right)-\partial_{y_{1}}\left(w_{0}u_{0z_{1}}\right)-\partial_{y_{1}}\left(w_{1}u_{0z_{0}}\right)-\partial_{y_{2}}\left(w_{0}u_{0z_{0}}\right)=\lambda_{\nu}^{-1}\left(\Omega_{2\xi_{1}\xi_{1}}+\Omega_{2y_{1}y_{1}}+\Omega_{4\eta_{0}\eta_{0}}\right)\]
%\end{comment}
%
We introduce the filtering over one wave period $\overline{\Omega}=\left(\sin\theta/2\pi\right)\int_{0}^{2\pi/\sin\theta}\mathrm{d}t\ \Omega$. This definition is slightly different from the filtering procedure in our experiment, but we keep the same notation. Filtering Eq. (\ref{eq:OmegaAD-1}) at order 5, and using the results obtained at order 0-4, we find that 
\begin{equation}
\Omega_{2t_{3}}=\left(\overline{\partial_{y_{1}}\left(w_{0}\partial_{z_1}u_{0}\right)}-\overline{\partial_{x_{1}}\left(w_{0}\partial_{z_0} v_{1}\right)}\right)+\lambda_{\nu}^{-1}\left(\Omega_{2\xi_{1}\xi_{1}}+\Omega_{2y_{1}y_{1}}\right)\ ,
\label{eq:omega_dyn_interm}
\end{equation}
with $\overline{\Omega}_{2}=\Omega_{2}$.%
Using Eq. (\ref{eq:sol_ordre1_W0_zeta1}-\ref{eq:v1-1}) as well
as $u_{0\eta_{0}}=w_{0\eta_{0}}/\tan\theta$, the Reynolds stresses
can be explicitely computed. For $t_{1}>\xi_{1}/\cos\theta$,
$t_{2}>\xi_{2}/\cos\theta$,  One finds that %
%\begin{comment}
%\begin{equation}
%\overline{\partial_{y_{1}}\left(w_{0}u_{0z_{1}}\right)}-\overline{\partial_{x_{1}}\left(w_{0}v_{1z_{0}}\right)}=\frac{-1}{\left(\sin\theta\right)^{2}}\frac{\partial_{y_{1}}\left(|\mathcal{A}_{0}|^{2}\right)}{\lambda_{\nu}}e^{-\xi_{1}/\left(\lambda_{\nu}\cos\theta\right)}\label{eq:order5_5}\end{equation} Finally, to avoid secular growth of $\Omega_{5}$ (using the fact that $\Omega_{1,2}$ does not depend on $t_{0}$), we must have 
%\begin{equation}
%\Omega_{2t_{3}}+\Omega_{4t_{1}}=\frac{-1}{\left(\sin\theta\right)^{2}}\frac{\partial_{y_{1}}\left(|\mathcal{A}_{0}|^{2}\right)}{\lambda_{\nu}}e^{-\xi_{1}/\left(\lambda_{\nu}\cos\theta\right)}\Theta\left(t_{1}-\frac{\xi_{1}}{\cos\theta}\right)+\frac{1}{\lambda_{\nu}}\left(\Omega_{2\xi_{1}\xi_{1}}+\Omega_{2y_{1}y_{1}}+\Omega_{4\eta_{0}\eta_{0}}\right)\label{eq:tout1}\end{equation}
%Given that $\Omega_{2}$ does not depend on $t_{1}$, given that $\Theta\left(t_{1}-\xi_{1}/\cos\theta\right)=1$ for sufficiently large $t_{1}$and given that $\Omega_{4}(t_{1}=0)=0$, we have necessarily $\Omega_{4t_{1}}=\Omega_{4\eta_{0}}=0$ and, for sufficiently large $t_{1}$,
%\end{comment}
\begin{equation}
\Omega_{2t_{3}}=\frac{-1}{\left(\sin\theta\right)^{2}}\frac{\partial_{y_{1}}\left(|\mathcal{A}_{0}|^{2}\right)}{\lambda_{\nu}}e^{-\xi_{1}/\left(\lambda_{\nu}\cos\theta\right)}+\frac{1}{\lambda_{\nu}}\left(\Omega_{2\xi_{1}\xi_{1}}+\Omega_{2y_{1}y_{1}}\right)
\label{eq:order5_6}
\end{equation}
%\begin{comment}
%Since $\mathcal{A}_{0t_{2}}=0$ for sufficiently large $t_{2}$ (see the end of the order 3 computation)
%\begin{equation}
%\Omega_{2t_{3}}=\frac{\partial_{x_{1}y_{1}}\left(|U_{0}|^{2}\right)}{\cos^{2}\theta}+\frac{1}{\lambda_{\nu}}\left(\Omega_{2\xi_{1}\xi_{1}}+\Omega_{2y_{1}y_{1}}\right)\label{eq:order5_7}\end{equation}
%Dans la version sur arxiv, on avait introduit $U^{*2}=4|U_{0}|^{2}$,
%ce qui donne
%\begin{equation}
%\Omega_{2t_{3}}=\frac{\partial_{x_{1}y_{1}}\left(U^{*2}\right)}{4\cos^{2}\theta}+\frac{1}{\lambda_{\nu}}\left(\Omega_{2\xi_{1}\xi_{1}}+\Omega_{2y_{1}y_{1}}\right)\label{eq:order5_8}\end{equation}
%Using $t_{3}=\epsilon^{3}Nt$, $U^{*}=U^{\prime}/U$, $x_{1}=\epsilon x/\lambda$,$%\xi_{1}=\epsilon\xi/\lambda$,$y_{1}=\epsilon y/\lambda$,
%$\Omega_{2}=\epsilon^{-2}\Omega\lambda/U,$ $\lambda_{\nu}=\epsilon L_{\nu}/\delt%a=\epsilon\lambda^{2}N/\nu$
%leads to 
%\begin{equation}
%N^{-1}\epsilon^{-5}\frac{\lambda}{U}\Omega_{t}=\frac{\lambda^{2}}{U^{2}}\epsilon^{-2}\frac{\partial_{xy}\left(U^{\prime2}\right)}{4\cos^{2}\theta}+\frac{\lambda^{2}}{\lambda^{2}}\frac{\nu}{N}\frac{\lambda}{U}\epsilon^{-3}\left(\Omega_{\xi\xi}+\Omega_{yy}\right)\label{eq:order5_9}\end{equation}
%Using $Fr_{\lambda}=\frac{U}{N\lambda}=\epsilon^{3}$ leads to 
%\end{comment}
%{}
Experimentally, we measure the wave amplitude of the velocity in the
$x$-direction $U^{\prime}=2|W_{0}|/\tan\theta$. Using Eq. (\ref{eq:sol_ordre1_W0_zeta1}-\ref{eq:v1-1}) and $u_{0\eta_{0}}=w_{0\eta_{0}}/\tan\theta$, $\mathcal{A}_{0}$ can be expressed in term of~$U^{\prime}$. Coming back to dimensional units we obtain the following equation for the temporally filtered vorticity
\begin{equation}
\overline{\Omega}_{t}=\frac{\partial_{xy}\left(U^{\prime2}\right)}{\left(2\cos\theta\right)^2}+\nu\Delta\overline{\Omega}\ .
\label{eq:order5_10}
\end{equation}
}


\begin{thebibliography}{0}%
\makeatletter
\providecommand \@ifxundefined [1]{%
 \@ifx{#1\undefined}
}%
\providecommand \@ifnum [1]{%
 \ifnum #1\expandafter \@firstoftwo
 \else \expandafter \@secondoftwo
 \fi
}%
\providecommand \@ifx [1]{%
 \ifx #1\expandafter \@firstoftwo
 \else \expandafter \@secondoftwo
 \fi
}%
\providecommand \natexlab [1]{#1}%
\providecommand \enquote  [1]{``#1''}%
\providecommand \bibnamefont  [1]{#1}%
\providecommand \bibfnamefont [1]{#1}%
\providecommand \citenamefont [1]{#1}%
\providecommand \href@noop [0]{\@secondoftwo}%
\providecommand \href [0]{\begingroup \@sanitize@url \@href}%
\providecommand \@href[1]{\@@startlink{#1}\@@href}%
\providecommand \@@href[1]{\endgroup#1\@@endlink}%
\providecommand \@sanitize@url [0]{\catcode `\\12\catcode `\$12\catcode
  `\&12\catcode `\#12\catcode `\^12\catcode `\_12\catcode `\%12\relax}%
\providecommand \@@startlink[1]{}%
\providecommand \@@endlink[0]{}%
\providecommand \url  [0]{\begingroup\@sanitize@url \@url }%
\providecommand \@url [1]{\endgroup\@href {#1}{\urlprefix }}%
\providecommand \urlprefix  [0]{URL }%
\providecommand \Eprint [0]{\href }%
\providecommand \doibase [0]{http://dx.doi.org/}%
\providecommand \selectlanguage [0]{\@gobble}%
\providecommand \bibinfo  [0]{\@secondoftwo}%
\providecommand \bibfield  [0]{\@secondoftwo}%
\providecommand \translation [1]{[#1]}%
\providecommand \BibitemOpen [0]{}%
\providecommand \bibitemStop [0]{}%
\providecommand \bibitemNoStop [0]{.\EOS\space}%
\providecommand \EOS [0]{\spacefactor3000\relax}%
\providecommand \BibitemShut  [1]{\csname bibitem#1\endcsname}%
\let\auto@bib@innerbib\@empty
%</preamble>
\end{thebibliography}%


\begin{thebibliography}{100}

\bibitem{Lighthill1978}
J.~Lighthill,  ``Waves In Fluids'', (Cambridge University Press, London, 1978).

\bibitem{Pedlosky1987}
J.~Pedlosky,  ``Geophysical Fluid Dynamics'',  (Springer-Verlag, Heidelberg, 1987).

\bibitem{Fincham00}
A.~Fincham, G.~Delerce, ``Advanced optimization of correlation imaging velocimetry algorithms'',
{Experiments in Fluids} {\bf 29}, 13 (2000).

\bibitem{Sutherland1999}
B.R. Sutherland, S.B. Dalziel, G.O. Hughes, and P.F. Linden,
 ``Visualization and measurement of internal waves by synthetic
  schlieren. part 1. vertically oscillating cylinder'',
 {Journal of Fluid Mechanics} {\bf 390}, 93--126, (1999).

\bibitem{Gostiaux2007}
L.~Gostiaux, H.~Didelle, S.~Mercier, and T.~Dauxois,
 ``A novel internal waves generator'',  {Experiments in Fluids} {\bf 42}, 123--130 (2007).

\bibitem{Mercier2010}
M.~Mercier, D.~Martinand, M.~Mathur, L.~Gostiaux, T.~Peacock, and T.~Dauxois,
 ``New wave generation'',  {Journal of Fluid Mechanics} {\bf 657}, 308--1334 (2010).

\bibitem{SutherlandBook}
B.R. Sutherland, ``Internal Gravity Waves'', (Cambridge University Press, London, 2010).

\bibitem{GostiauxDauxois2007}
L.~Gostiaux, T.~Dauxois,
 ``Laboratory experiments on the generation of internal tidal beams over
  steep slopes'',
 {Physics of Fluids} {\bf 19}, 028102 (2007).
 
\bibitem{ManiTomPRL}
M.~Mathur, T.~Peacock, ``Internal wave interferometry'',
 {Physical Review Letters} {\bf 104}, 118501 (2010).

\bibitem{Peacock2009}
T.~Peacock, M.J. Mercier, H.~Didelle, S.~Viboud, and T.~Dauxois,
 ``A laboratory study of low-mode internal tide scattering by
  finite-amplitude topography'',
 {Physics of Fluids} {\bf 21}, 121702 (2009).

\bibitem{Joubaud2011}
S.~Joubaud, J.~Munroe, P.~Odier, and T.~Dauxois,
"Experimental parametric subharmonic Instability in stratified fluids,"
 {Physics of Fluids}  {\bf 24}, 041703 (2012).

\bibitem{VallisBook}
G.~K. Vallis,  {``Atmospheric and Oceanic Fluid Dynamics''},
 (Cambridge University Press, London, 2006).

\bibitem{BuhlerBook}
O.~K. Buhler,  {``Waves and Mean Flows''},
  (Cambridge University Press, London, 2009).


\bibitem{DauxoisYoung}
{T. Dauxois, W. R. Young, ``{Near critical reflection of internal waves}'', {Journal of Fluid Mechanics} {\bf 390}, 271-295 (1999).}

\bibitem{Akylas}
{A. Tabaei, T. R.  Akylas, ``{Nonlinear internal gravity wave beams}'', {Journal of Fluid Mechanics} {\bf 482}, 141-161 (2003).}

\bibitem{king}
{B. King, H.P.  Zhang, H.L. Swinney, ``{Tidal flow over three-dimensional topography in a stratified fluid}'', {Physics of Fluids} {\bf 21}, 116601 (2009).}

\bibitem{Bretherton1969}
F.~P. {Bretherton},  ``{On the mean motion induced by internal gravity waves}'',
 {Journal of Fluid Mechanics} {\bf 36}, 785--803 (1969).

\bibitem{Lelong1991}
M.-P. Lelong, J.~Riley,
 ``Internal wave-vortical mode interactions in strongly stratified
  flows'',
 {Journal of Fluid Mechanics} {\bf 232}, 1--19 (1991).

%\bibitem{uvmat} J.~Sommeria,  LEGI / CNRS-UJF-INPG, {\url{http://coriolis.legi.grenoble-inp.fr}}.

\bibitem{Mercier2008}
M.~Mercier, N.~Garnier,  T.~Dauxois,
 ``Reflection and diffraction of internal waves analyzed with the
  Hilbert transform'',
 {Physics of Fluids} {\bf 20}, 0866015 (2008).

\bibitem{AkylasB}
{A. Tabaei, T. R.  Akylas, K. G. Lamb, ``{Nonlinear effects in reflecting and colliding internal wave beams}'', {Journal of Fluid Mechanics} {\bf 526}, 217-243 (2005).}

\bibitem{Leclair2011}
%M.~Leclair, N.~Grisouard, L.~Gostiaux, C.~Staquet, F.~Auclair,
 %``Reflexion of a plane wave onto a slope and wave-induced mean flow'', {International Symposium of Stratified Fluids}, Roma (2011).
N.~Grisouard,  ``R\'eflexions et r\'efractions non-lin\'eaires d'ondes de gravit\'e internes'', 
PhD Thesis, Universit\'e de Grenoble, 2010 (http://tel.archives-ouvertes.fr/tel-00540608/en/).
 (http://tel.archives-ouvertes.fr/tel-00540608/en/). 
M. Leclair, N. Grisouard, L. Gostiaux, C. Staquet, F. Auclair,
"Reflexion of a plane wave onto a slope and wave-induced mean flow",
Proceedings of the VII International Symposium on stratified flows,
Rome 22-26 August 2011, Editor: Sapienza Universit\`a di Roma,  ISBN: 9788895814490.


\bibitem{Grisouard2012}
N.~Grisouard, O.~Buhler,
 ``Forcing of oceanic mean flows by dissipating internal tides'', {in press, Journal of Fluid Mechanics} (2012).

\end{thebibliography}
\end{document}